# Tribute to Henry Primakoff: Chiral Perturbation Theory Tests via Primakoff Reactions

**Murray Moinester***

School of Physics and Astronomy, Tel Aviv University, 69978 Tel Aviv, Israel

*e-mail: murray.moinester@gmail.com

**Abstract.** Consider high-energy (GeV) beam particles scattering from the Coulomb field of a target nucleus (Z, A). The Coulomb field acts as a target of $\gamma^*$ virtual photons, with the target density proportional to $Z^2$. Henry Primakoff was the first to propose determining the lifetime of the $\pi^0$ meson by measuring the production cross section for the reaction $\gamma\gamma^* \rightarrow \pi^0$. This process occurs when a high-energy gamma-ray interacts with the Coulomb field. Quasi-real exchanged photons ($\gamma^*$) are identified by isolating the sharp Coulomb peak at very low values of the squared four-momentum transfer t to the target nucleus. The scattering cross section via one-photon exchange is proportional to the fine-structure constant α and inversely proportional to $t^2$. Since t is inversely related to the squared center-of-mass energy (s), it decreases rapidly as s increases. Consequently, despite the weakness of the electromagnetic interaction, the interaction amplitude can still be significant. We will first discuss Primakoff's scientific career and personal life. Next, we will review the Primakoff scattering experiments that measured the pion polarizability and the $\gamma \rightarrow \pi\pi\pi$ chiral anomaly amplitude at CERN COMPASS and the $\pi^0$ lifetime at Jefferson Laboratory (JLab). The data from these experiments are in good agreement with two-flavor (u, d) Chiral Perturbation Theory (ChPT) predictions. We explain that additional Primakoff measurements with kaons and η mesons are needed to test how well three-flavor (u, d, s) ChPT captures strange-quark effects.

## 1. Introduction

Henry Primakoff helped found a field of study based on the investigation of two-body scattering of very high energy particles (pions, kaons, gamma rays) from the Coulomb field of nuclei. In this kinematic regime, the squared four-momentum transfer t to the target nucleus decreases rapidly with increasing collision energy, which reduces the relative contribution of strong interaction processes. When a high-energy (190 GeV/c) pion beam is used, Primakoff scattering, viewed in the rest frame of the pion, is effectively low-energy soft scattering of ≈250 MeV gamma rays from a pion at rest. We describe Primakoff scattering studies of pion polarizability, the $\gamma \rightarrow \pi\pi\pi$ chiral anomaly and the $\pi^0$ lifetime. Following a review of Chiral Perturbation Theory (ChPT), we discuss the good agreement of these studies with 2-flavor (u, d) ChPT predictions. We discuss why future Primakoff studies at CERN AMBER and Jefferson Laboratory (JLab) will be important for testing the theoretical framework of 3-flavor (u, d, s) ChPT. These studies include kaon polarizability, $\gamma \rightarrow \pi\pi\eta$ and $\gamma \rightarrow KK\pi$ chiral anomalies, and η lifetime. The transition from 2-flavor to 3-flavor ChPT incorporates a strange quark, which is crucial for understanding the full dynamics of light mesons (pions, kaons, etas). Comparing 3-flavor ChPT predictions with data for kaons, $\pi^0$ and η will allow us to assess how well ChPT captures the effects of strange quarks.

As informal spokesman of the CERN COMPASS Primakoff program, I was responsible for writing the Primakoff section of the proposal in 1996 [1,2]. COMPASS physics experiments began in 2002 with a muon beam and polarized proton and deuteron targets [3–5]. Pion polarizability, chiral anomaly, and radiative transition measurements began in 2009 using Primakoff scattering of pions from high-Z nuclei. CERN AMBER phase-1 [6,7], which began in 2022, is investigating fundamental questions regarding the origin of hadronic mass in the universe. AMBER phase-2 plans to study kaon-induced Primakoff reactions, following a COMPASS (M2) beam line upgrade that sets up radio-frequency-separated high-energy and high-intensity kaon and antiproton beams [8]. I recently presented a brief review of Primakoff tests of three-flavor ChPT [9].

## 2. Henry Primakoff Biography

Henry Primakoff [10], 1914–1983, was the first Donner Professor of Physics at U. Penn. He graduated from Columbia University in 1936 and obtained his Ph.D. in Physics from New York University in 1938. He was a theoretical physicist well known for his contributions to condensed matter and high-energy physics. He helped develop the Holstein–Primakoff transformation [11], a mapping from boson creation and annihilation operators to spin operators, whereby spin waves in ferromagnets are treated as bosonic excitations. Primakoff became a leading authority in weak interaction phenomena by formulating a muon-nucleon effective Hamiltonian. Fujii and Primakoff [12,13] calculated the partial muon capture rates in certain light nuclei, which were in good agreement with the experimental results. The Fujii-Primakoff Hamiltonian was also used to describe recoil nuclear polarization in muon capture [14]. Primakoff also contributed to the understanding of double beta decay [15,16], neutrino-nucleus scattering [17], and shock waves in water [18].

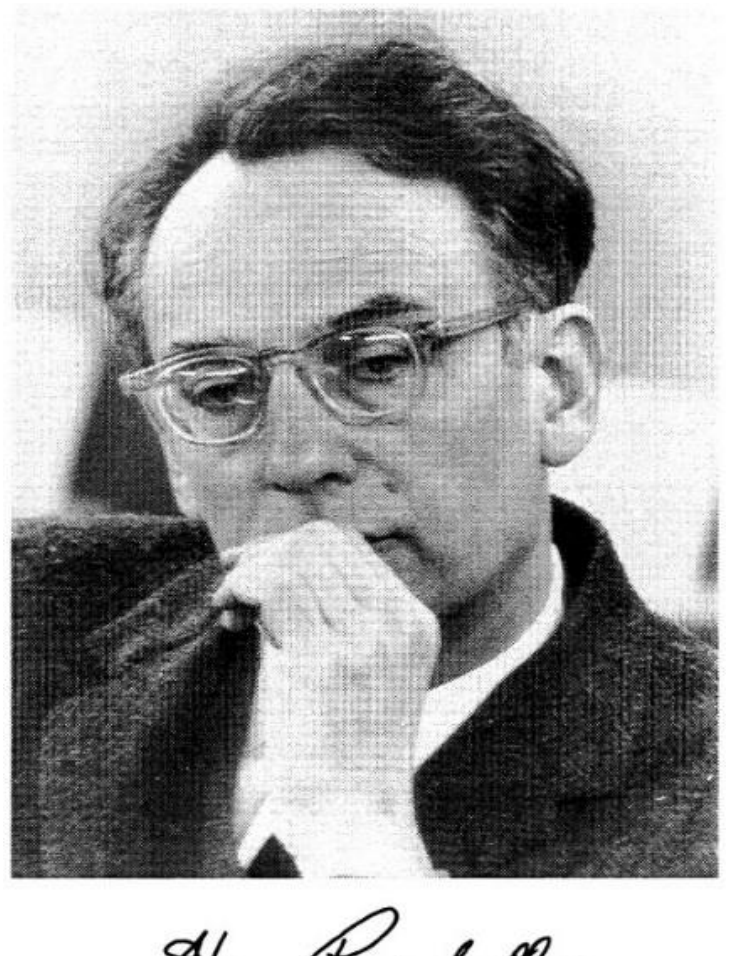

**Figure 1.** Henry Primakoff photo, shown courtesy of the University of Pennsylvania.

During his university studies, he met biochemist Mildred Cohn, 1913-2009, whom he married in 1938, who pioneered the use of NMR to study enzyme reactions, and who also became a full professor at U. Penn. The couple had three children. Figure 1 shows a photograph of Henry Primakoff, courtesy of the University of Pennsylvania.

Through his mother, Henry descended from a large Jewish merchant family that had lived in Odessa for several generations. Through his father, Henry came from a wealthy and prestigious Greek Orthodox family. Henry's father was born in Kiev, where he studied medicine, and graduated as a doctor in 1911. His mother came from Odessa to Kiev to study pharmacy, and it was through their medical connections that they met. During WWI, his father served as an army doctor, was wounded while operating on soldiers, and died in 1919 shortly after WWI ended and the Russian civil war began. Henry's family decided to leave Odessa. This required escaping across the Prut River into Romania, traveling for long hours through the woods at night, hiding during the day in remote farmhouses, and ultimately finding refuge on the farm of some relatives. Henry was not allowed to talk when they went into town, because it was too dangerous to speak Russian. The family successfully obtained Romanian travel documents and embarked on a lengthy journey through war-torn Europe to Bremen, followed by a steamship voyage to New York, where they settled in 1922.

## 3. Primakoff Effect and Primakoff Scattering

The Primakoff Effect [19], $\gamma\gamma^* \to \pi^0$, refers to the resonant production of neutral pseudoscalar mesons ($\pi^0$ or $\eta$) through the two-body interaction of high-energy photons with quasi-real photons $\gamma^*$ in the nuclear Coulomb field. This is equivalent to the inverse kinematics process of $\pi^0$ decaying into two photons, and has been utilized to measure the decay width (lifetime) of neutral mesons. In the related Primakoff scattering [20], high-momentum beam particles (pions, kaons, or gamma rays), such as 190 GeV/c $\pi^-$, scatter from virtual photon targets $\gamma^*$ within the nuclear Coulomb field, while the target nucleus remains in its ground state. This process is known as Bremstrahlung, when high-energy electrons are incident on a target. This can be conceptualized as the Compton scattering of electrons from virtual photons. The scattered electrons and real photons are projected into forward angles in the laboratory frame. For any beam particle, the virtual photons serve as targets whose effective thickness is advantageously proportional to $Z^2$. Although a Pb (Z=82) target would maximize event rates, a medium-Z target such as Ni (Z=28) is preferred to reduce systematic errors from double-photon exchange and nuclear backgrounds. The Weizsäcker-Williams equivalent photon approximation [21,22] formalizes this framework, enabling the extraction of low-energy quantities from the data. Fermi first suggested this approximation, noting that the electromagnetic field of a rapidly moving charged particle is equivalent to a photon pulse [23]. Here, the spherically symmetric Coulomb electromagnetic field of the target nucleus manifests as a transverse photon pulse when viewed from the rest frame of the incident beam particle. One-photon exchange scattering is proportional to the fine-structure constant $\alpha$ and inversely proportional to the squared four-momentum transfer t. The value of t is inversely proportional to the squared four-momentum center-of-mass energy s, and decreases rapidly as s increases. Despite the weak electromagnetic interaction, the interaction amplitude can therefore still be significant. In this quasi-elastic process, the contribution of one photon exchange is enhanced, and the very small value of t compensates for the weak electromagnetic coupling. Moreover, the small t value allows for a clear distinction between ultra-peripheral Coulomb scattering and nuclear background interactions involving larger momentum transfers. In addition, the nuclear form factor at very low t is at its maximum; therefore, it does not suppress the scattering amplitude.

## 4. The ChPT effective Lagrangian $L_{eff}$

The Quantum Chromodynamics (QCD) Lagrangian $L_{QCD}$ encodes the strong force, including color charge and quark-gluon dynamics [24]. As $m_q \to 0$, left- and right-handed fields $q_L$ and $q_R$ decouple, and $L_{QCD}$ acquires an $SU(3)_L \times SU(3)_R$ chiral symmetry. Chirality refers to the handedness defined by the relative orientation of spin and momentum. The index q denotes the different quark flavors u, d, and s. However, chiral symmetry is explicitly broken by the small, nonzero quark masses; and spontaneously broken by the QCD vacuum, which does not respect the symmetry. Consequently, the $L_{QCD}$ symmetry is reduced to the diagonal subgroup $SU(3)_{L+R}$. At very short distances ($\lesssim$0.1 fm), equivalently, high energies, the QCD running coupling constant becomes small (asymptotic freedom); therefore, quarks and gluons interact weakly and perturbation theory is applicable. At larger distances, 0.5–1 fm, the coupling becomes large, perturbation theory breaks down and non-perturbative phenomena (confinement, chiral symmetry breaking) dominate. The 0.1–0.5 fm range is a crossover regime: perturbation theory becomes increasingly marginal while nonperturbative effects grow, so accurate descriptions require hybrid or nonperturbative approaches (lattice QCD, effective models such as ChPT, and resummation techniques).

Low-momentum transfer calculations can be carried out within the effective field theory introduced by Weinberg [25] and developed by Gasser & Leutwyler [26]. Chiral perturbation theory (ChPT) is a systematic low-energy expansion of QCD implemented by an effective Lagrangian $L_{eff}$ written as a power series in meson momenta and mass terms [27–30]. $L_{eff}$ encodes the dynamics of the light, low-energy degrees of freedom (the pseudo-Goldstone bosons) that arise from spontaneous chiral symmetry breaking, while high-energy quark-gluon dynamics are integrated out. Goldstone's theorem asserts that each spontaneously broken continuous symmetry is associated with a massless scalar particle; in QCD the approximate chiral symmetry produces light pseudo-Goldstone mesons. Restricting to u,d flavors yields the pions ($\pi^+$, $\pi^-$, $\pi^0$) as the pseudo-Goldstone modes; including s yields the light octet $J^P = 0^-$ mesons ($\pi^0$, $\pi^\pm$, $K^\pm$, $K^0$, $\overline{K}^0$, $\eta$). Consequently $L_{eff}$ (ChPT) is expected to provide a good description of pion, kaon, and eta interactions at low energies and small momentum transfers.

Gell-Mann, Oakes and Renner [31] and Gasser & Leutwyler [32] formally described how a "chiral condensate" manifests in the masses of Goldstone bosons. The QCD vacuum is not empty; it has a highly nontrivial structure characterized by a nonzero quark condensate, which serves as an order parameter for spontaneous chiral symmetry breaking [24]. The chiral condensate quantifies the vacuum expectation value of quark-antiquark pairs, $\langle 0|\bar{q}\, q|0\rangle$, which reflects the degree of vacuum polarization. Ref. 31 explicitly demonstrated that the squared pion mass is proportional to the average bare mass of the light quarks (u, d) times the magnitude of the chiral condensate. This condensate is responsible for generating constituent quark masses and accounts for a significant portion of hadron masses. The agreement between the observed pion mass and this relationship supports the notion that the magnitude of vacuum condensate fundamentally governs the pattern of chiral symmetry breaking. This connection between the vacuum structure and low-energy observable properties provides a robust theoretical foundation and justifies the use of ChPT, which exploits the spontaneously broken symmetry and effective Lagrangian framework to describe mesonic phenomena at low energies.

Note that while ChPT uses physical meson masses as inputs and observables, the formal expansion is controlled by momenta and the light current quark masses ($m_q/\Lambda_\chi$), where the chiral scale is $\Lambda_\chi \approx 1$ GeV. The estimated quark masses ($m_u \approx 2$ MeV, $m_d \approx 5$ MeV, $m_s \approx 93$ MeV) are $\overline{MS}$ (MS-bar) modified minimal subtraction values evaluated at the renormalization scale $\mu = 2$ GeV [33]. At leading order, meson squared masses scale with $m_q$ times the quark condensate, so kaons and the $\eta$ are also formally pseudo-Goldstone bosons. However, SU(3) expansions may converge more slowly or irregularly or less reliably than SU(2), because the strange-quark mass is larger compared with the chiral scale. There is no guarantee that adding higher orders will fix a large leading-order discrepancy, since higher-order terms can be large or poorly constrained. Reliable application to K/$\eta$ processes therefore requires explicit order-by-order convergence checks, well-determined Low Energy Constants or lattice QCD input, and/or alternative theoretical strategies. The first order of business is to achieve experimental data.

In this discussion, we focus on processes involving photons and mesons. Photons are incorporated into $L_{eff}$ through the introduction of electromagnetic coupling terms that respect chiral symmetry. ChPT then provides rigorous predictions for $\gamma\pi$ interactions at low energies. Unitarity is restored by including pion-loop corrections, whose ultraviolet divergences are absorbed into renormalized low-energy constants $L_r$. By restricting the perturbative expansion of $L_{eff}$ to quartic order in momenta and masses, $O(p^4)$, this approach establishes connections between various processes through a common set of 12 $L_r$ constants, which encapsulate the influence of perturbative QCD physics within an effective framework [26]. The pion polarizabilities discussed below are linked (via two of the coupling constants) to the ratio $h_A/h_V$ measured in radiative pion beta decay $\pi^+ \to e^+\nu_e\gamma$, where $h_A$ and $h_V$ are the axial vector and vector coupling constants of the decay [34]. ChPT is applicable to the Primakoff reactions related to polarizabilities, chiral anomalies, and meson lifetimes currently under investigation at CERN and JLab.

## 5. Gamma-Pion Compton scattering and Pion Polarizabilities

Polarizabilities have long been associated with the scattering cross-section of sunlight photons interacting with molecular electrons in atmospheric $N_2$ and $O_2$. At optical wavelengths, the incident photon energies ($\approx$1.6 – 3.2 eV) are small compared with the typical electronic binding energies, which are in the tens of eV range. The oscillating electric field of sunlight photons induces vibrations of molecular electrons. Associated with the varying electric dipole moment, energy is radiated in proportion to the square of the second time derivative (acceleration) of the dipole moment. The radiated power is given by Power $\approx \alpha_a^2\lambda^{-4}$, where $\alpha_a$ is the electric polarizability of the molecule. Consequently, the scattering cross section depends on $\lambda^{-4}$, which means that blue light is scattered much more than red light.

Consequently, the intensities of the scattered and transmitted sunlight are dominated by blue and red light, respectively. This is why the daytime sky is blue and sunrises and sunsets are red. This phenomenon, known as Rayleigh scattering, is named after Lord Rayleigh [35].

The $\gamma\pi \rightarrow \gamma\pi$ Compton scattering cross section depends predominantly on the pion charge. In the pion rest frame, for γ energies in the range of $\omega = 60 - 780$ MeV and scattering angles greater than 80°, ≈5% of the differential cross section depends on the electric $\alpha_\pi$ and magnetic $\beta_\pi$ charged pion polarizabilities. These quantities describe the induced dipole moments of the pion during scattering. The electromagnetic field of the photon interacts with the quark substructure of the pion, inducing a variable electric dipole moment proportional to the electric field with proportionality constant $\alpha_\pi$, and a variable magnetic dipole moment proportional to the magnetic field with proportionality constant $\beta_\pi$. Polarizabilities are fundamental hadron characteristics [34,36,37]. The experimental ratio for $h_A/h_V$ (discussed above) leads to $\alpha_\pi - \beta_\pi = 5.4 \times 10^{-4}$ fm$^3$ at lowest order, and $\alpha_\pi - \beta_\pi = (5.7 \pm 1.0) \times 10^{-4}$ fm$^3$ [38] in a higher order two-loop approximation.

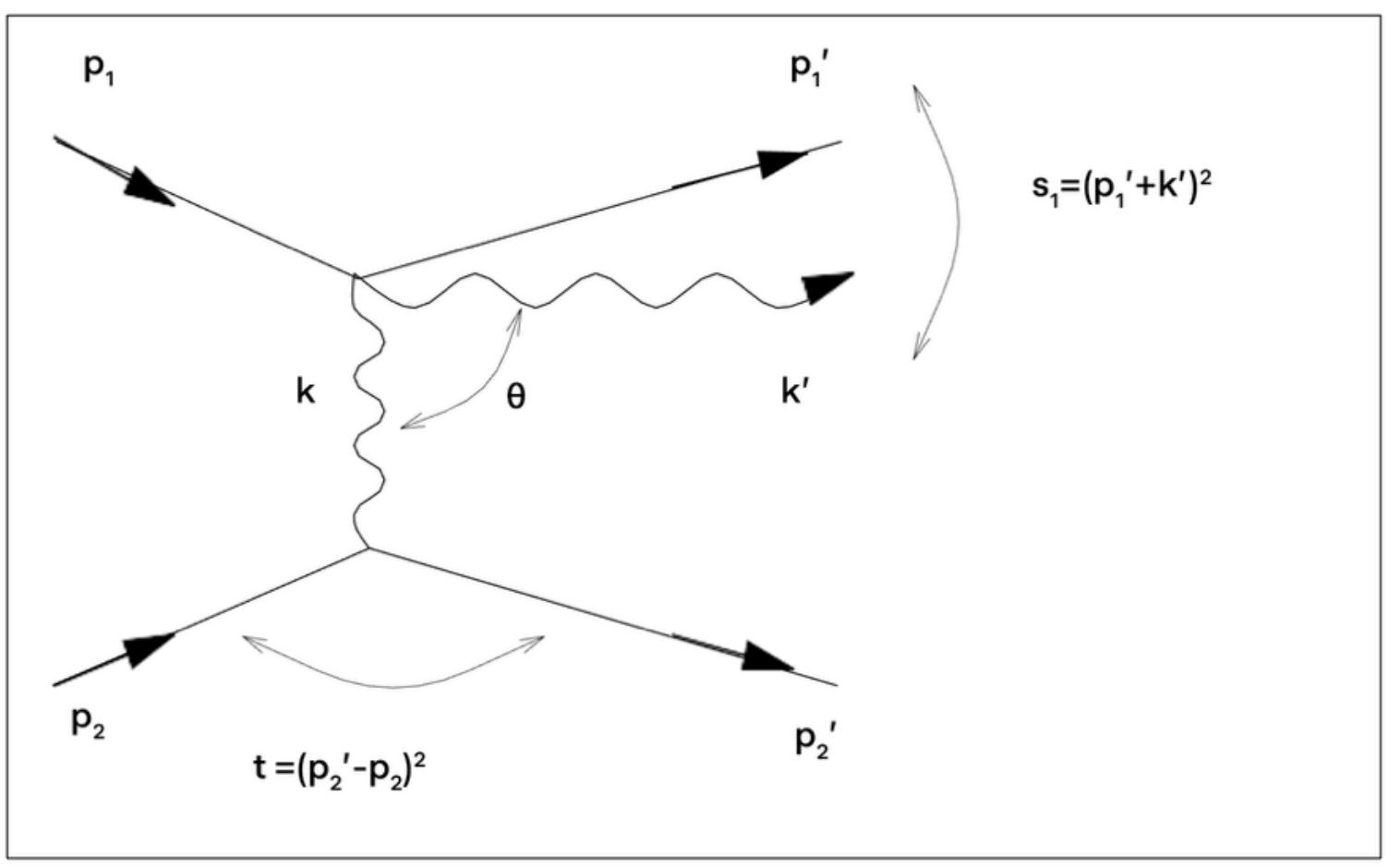

**Figure 2.** Primakoff γπ Compton process and kinematics. Four-momenta: $p_1$ (incoming pion), $p_{1'}$ (outgoing pion), $p_2$ (initial nucleus), $p_{2'}$ (recoiling nucleus), k (incoming virtual photon) and k′ (outgoing real photon). θ is the photon scattering angle.

The polarizabilities can be extracted from the shape of the γπ Compton scattering differential cross sections [34,39]. The influence of polarizability increases with increasing γ energy and γ scattering angle in the pion rest frame. COMPASS measured γπ scattering with 190 GeV/c negative pions through Primakoff scattering, $\pi Z \rightarrow \pi Z\gamma$, with a Ni (Z = 28) target [40]. In the one-photon exchange (Primakoff) regime, this process is equivalent to $\pi\gamma \rightarrow \pi\gamma$ scattering. Figure 2 illustrates the kinematic variables: $p_1$ and $p_{1'}$ for the initial and final pions, $p_2$ and $p_{2'}$ for the initial and final target nucleus Z, and k and k′ for the initial and final photons. Let $p_1$ denote the incident pion momentum in the laboratory. A virtual photon with 4-momentum $k = \{\omega, \vec{k}\} = p_2 - p_{2'}$ scatters from that pion; the squared 4-momentum transfer to the target nucleus is $t = M^2 = k^2 = \omega^2 - |\vec{k}|^2$. Because $t = 2M_Z\,[M_Z - E_Z(\text{lab})] < 0$ for an intact recoiling nucleus, the virtual photon mass M is imaginary. Because t is near zero (Coulomb peak), the virtual photon is nearly transverse and on-shell, so pion Compton scattering off that photon is a good approximation to real-photon Compton scattering. The virtual photon momentum $\vec{k}$ is transverse to the beam, and is equal in magnitude and opposite in direction to the transverse recoil momentum $p_T$ of the nucleus. In the pion rest frame, θ denotes the photon scattering angle relative to the incident virtual-photon direction. Scattered photons (γ) and pions emerge at very high energies at forward laboratory angles. The squared 4-momentum of the γπ final state is $s_1$; therefore, the final state invariant mass is $m_{\pi\gamma} = \sqrt{s_1}$. The exchanged quasi-real photons are selected by isolating the sharp Coulomb peak observed at the lowest squared 4-momentum transfers to the target nucleus, denoted by t or $Q^2$. In COMPASS, the typical minimum value of the negative squared 4-momenta transfer was $Q^2_{min} = (1 \text{ MeV/c})^2$, and $Q^2 < 0.0015 \text{ GeV}^2/c^2$ was required.

In the pion rest frame, the virtual-photon energy is approximately $\omega \approx (s_1 - m_\pi^2)/2m_\pi$. The software cuts on $s_1$ were defined by selecting $m_{\pi\gamma}$ between 190 and 490 MeV/c², which corresponds to effective photon energies ω = 60−780 MeV, with ω ≈ 250 MeV being the approximate average energy. An excellent resolution in t is crucial, as the characteristic signature of Primakoff scattering involves very low t. This necessitates high angular resolution for the final state pion, achieved by using thin targets and detectors to minimize multiple Coulomb scatterings. The COMPASS analysis required exactly one final-state photon and one charged particle, whose combined four-momentum equals that of the incident beam. Pion Primakoff scattering is an ultra-peripheral reaction on a virtual photon target. The scattered pions have impact parameter b ≈ 50 fm with respect to the nucleus, minimizing nuclear-interaction backgrounds. This arises from the four-momentum transfer Q to the target nucleus, which extends up to $3Q_{min}$ and has an average value of approximately $2Q_{min}$. By the uncertainty principle, with $Q_{min} \approx 1$ MeV/c and $\Delta Q_{min} \approx 2$ MeV/c, the impact parameter is $\Delta b \approx \hbar c/(2c\Delta Q_{min}) \approx 197.3/4 \approx 50$ fm.

Assuming $\alpha_\pi + \beta_\pi = 0$ [34], the dependence of the differential cross-section on $x_\gamma = E_\gamma/E_\pi$ determines $\alpha_\pi$. Here $x_\gamma$ is the fraction of the beam energy carried by the final state γ. This variable is related to the $\gamma\pi \to \gamma\pi$ photon scattering angle. The chosen $x_\gamma$ range (0.4 − 0.9) corresponds to photon scattering between 80° and 180° in the pion rest frame, where sensitivity to $\alpha_\pi - \beta_\pi$ is maximal. Let $\sigma_E(x_\gamma, \alpha_\pi)$ and $\sigma_{MC}(x_\gamma, \alpha_\pi)$ be respectively the experimental and Monte Carlo calculated lab-frame differential cross-section as a function of $x_\gamma$ for a pion having polarizability $\alpha_\pi$. The term $\sigma_{MC}(x_\gamma, \alpha_\pi = 0)$ is the cross-section for a zero polarizability pion. The $\sigma_E(x_\gamma, \alpha_\pi)$ values were obtained by subtracting the background from the $\pi^-\text{Ni} \to \pi^-\text{Ni}\gamma$ diffractive channel, and from $\pi^-\text{ Ni} \to \pi^-\ \pi^0$ Ni diffractive and Primakoff channels. The experimental ratios are $R_\pi = \sigma_E(x_\gamma, \alpha_\pi)/\sigma_{MC}(x_\gamma, \alpha_\pi = 0)$. The polarizability $\alpha_\pi$ and its statistical error were obtained by fitting $R_\pi$ to the theoretical expression $R_\pi = 1 - 10^{-4} \times 72.73\, x_\gamma^2\, \alpha_\pi/(1-x_\gamma)$ [34], where $\alpha_\pi$ is in units of $10^{-4}$ fm³. The best-fit theoretical ratio $R_\pi$ is depicted in Figure 3 as the solid curve [41]. Systematic uncertainties were controlled by measuring $\mu\text{Ni} \to \mu\text{Ni}\gamma$ cross-sections. The main systematic uncertainties arise from the Monte Carlo simulations of the COMPASS setup. By comparing the measured and theoretical $x_\gamma$ dependences of $R_\pi$, and assuming $\alpha_\pi = -\beta_\pi$, one obtains $\alpha_\pi - \beta_\pi = (4.0 \pm 1.2_{stat} \pm 1.4_{syst}) \times 10^{-4}$ fm³ [41]. The good agreement with ChPT supports the identification of pions as Goldstone bosons.

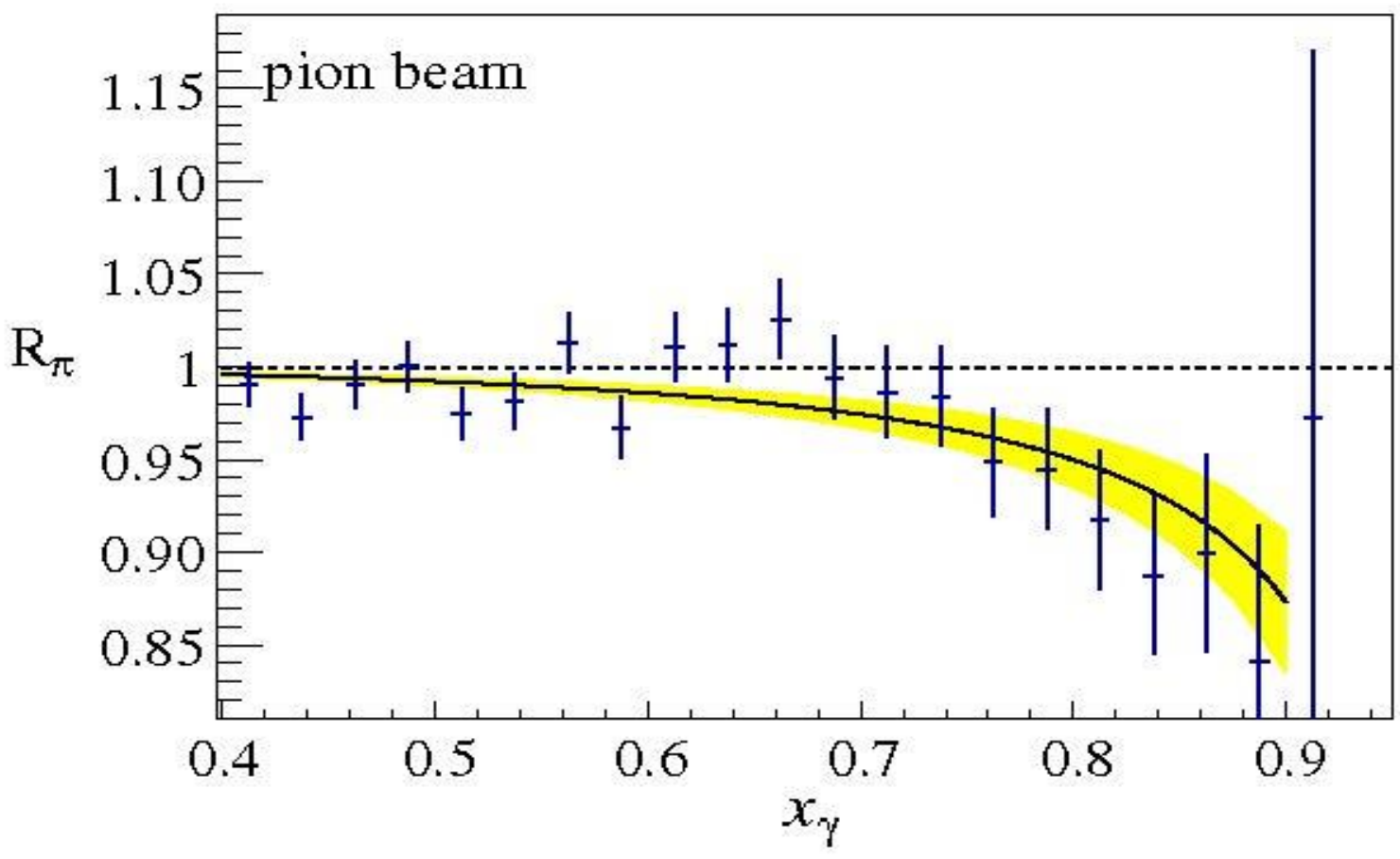

**Figure 3.** Determination of the pion polarizability by fitting the experimental $R_\pi$ $x_\gamma$ distribution (data points) to the theoretical Monte Carlo $R_\pi$ (solid line), from Ref. [41]. The yellow band denotes the statistical uncertainty of the fit shown by the solid line. Error bars denote statistical uncertainties.

## 6. Future Polarizability Studies

**Jlab:** Proposal E12-13-008 [42] plans to measure $\gamma\gamma \to \pi^+\pi^-$ cross-sections and asymmetries via the Primakoff reaction utilizing a 6 GeV linearly polarized photon beam, a Sn "photon target" and the JLab GlueX detector. The azimuthal angle asymmetry dependence will be used to reduce the backgrounds. The goal is to achieve an uncertainty of approximately 10% for $\alpha_\pi - \beta_\pi$.
**COMPASS & AMBER:** Higher statistical data (≈5X) already collected by COMPASS are expected to provide an improved determination of $\alpha_\pi - \beta_\pi$. Kaon polarizabilities have been proposed to be measured at CERN AMBER using an RF-separated kaon beam.

## 7. ChPT for $\gamma \to 4\pi$

COMPASS measured $\pi^-\gamma \to \pi^-\pi^-\pi^+$ via the Primakoff reaction $\pi^-Pb \to \pi^-\pi^-\pi^+Pb$ in the low-mass region of $m_{3\pi} < 5m_\pi$, well below the $a_1(1260)$ and $a_2(1670)$ resonances [40]. The cross section data were measured with a total uncertainty of ≈20%. These data are in good agreement with the lowest-order ChPT cross-section predictions.

## 8. Chiral Axial Anomaly

Classically the axial (chiral) current is conserved, but quantum effects (e.g., the fermion triangle diagram) produce the Adler–Bell–Jackiw anomaly, so that the divergence of the axial current is anomalously nonzero. For $\gamma\pi$ interactions, the low-energy effective Lagrangian $L_{eff}$ must therefore include the Wess–Zumino–Witten (WZW) functional [43–45], in order to reproduce QCD's anomalous Ward identities. The WZW coefficient is quantized (proportional to the number of colors $N_c$), reflecting its topological origin. The WZW action, written in terms of the pseudoscalar Goldstone octet ($\pi$, K, $\eta$), appears at $O(p^4)$ in chiral counting. It governs the odd-intrinsic-parity sector, and generates vertices that involve one or more photons coupled to an odd number of Goldstone bosons. The chiral anomaly therefore controls processes such as $\pi^0 \to \gamma\gamma$, $\gamma \to 3\pi$ (equivalently $\gamma\pi \to \pi\pi$), $\gamma\pi \to \pi\eta$ [46], and $\gamma K \to K\pi^0$ [47]. At leading order, these amplitudes (e.g., $F_\pi$ for $\pi^0 \to 2\gamma$ and $F_{3\pi}$ for $\gamma \to 3\pi$) are fixed by the WZW term; higher-order chiral loops and local $O(p^6)$ counter terms provide calculable corrections.

## 9. The $\pi^0 \to \gamma\gamma$ $F_\pi$ Amplitude

The chiral anomaly amplitude $F_\pi$ provides an important measure of the interaction strength of pions and directly influences the decay width $\Gamma$ and the mean lifetime $\tau$ of the $\pi^0$. These two quantities are related through the Heisenberg Uncertainty Principle, expressed as $2(\Delta E)(\Delta t) = \hbar$, or $\Gamma\tau = \hbar$, where $\Gamma = 2\Delta E$ is the FWHM of the resonance peak in the experimentally measured $\pi^0$ mass distribution, and $\Delta t$ is the total time (lifetime $\tau$) available for the mass measurement. Thus, we have $\Gamma\tau = 65.82 \times 10^{-17}$, where $\Gamma$ is measured in eV and $\tau$ is measured in seconds. Since $\Gamma(\pi^0 \to \gamma\gamma) = BR(\pi^0 \to \gamma\gamma)\Gamma(\pi^0)$, where the Branching Ratio is BR = 0.988, the full $\pi^0$ lifetime is calculated using $\Gamma(\pi^0)$ rather than $\Gamma(\pi^0 \to \gamma\gamma)$.

ChPT at $O(p^4)$ predicts $F_\pi = \alpha/\pi f = 0.0252$ GeV$^{-1}$, where $\alpha$ is the fine structure constant and f is the pion decay constant [48,49]. The decay width is given [50] by ChPT as $\Gamma(\pi^0) = \alpha^2 m^3/64\pi^3 f^2$ eV, and the lifetime in seconds is $\tau = 65.82 \times 10^{-17}/\Gamma$. The calculations and experimental results were reviewed by Bernstein and Holstein [50] and Miskimen [51]. Table 1 summarizes the results for $\pi^0 \to 2\gamma$ decay, combining the statistical and systematic errors in quadrature when both are provided. The most recent and precise Primakoff effect measurement of the $\pi^0$ lifetime was carried out at JLab by the PrimEx collaboration (named after Primakoff), with the result $\tau(\pi^0) = (8.34 \pm 0.06 \text{ stat.} \pm 0.11 \text{ syst.}) \times 10^{-17}$ s [52].

However, the PrimEx lifetime result significantly disagrees with the direct lifetime method [53], which is based on the $\pi^0 \to 2\gamma$ mean decay distance. The direct measurement employed forward angle (≈0°) $\pi^0$ mesons produced by 450 GeV/c protons incident on a target consisting of two tungsten foils, arranged to allow variable separations. The $\pi^0$ decays were observed as a function of separation by detecting 150 GeV/c positrons produced by decay γ-rays that converted in the foils. The average $\pi^0$ momentum (150 – 410 GeV/c) that produced such positrons was estimated to be $\langle P(\pi^0)\rangle$ = 235 GeV/c, which was used in the lifetime calculation. However, the required $\pi^0$ momentum spectrum was not measured; instead, it was estimated by averaging the momenta spectra of $\pi^+$ and $\pi^-$, which were measured at 150-300 GeV/c and estimated for 300–410 GeV/c. One possible reason for this disagreement is that the $\tau(\pi^0)$ error by this method may be larger than the estimated 3% uncertainty [50,54]. Regarding the $e^+e^- \to e^+e^-\gamma\,\gamma$ result [55], the 9% uncertainty is too large to rigorously test the theoretical calculations. Therefore, we focus on the PrimEx result.

**Table 1.** Theoretical and experimental results for the $\pi^0 \to 2\gamma$ decay.

| Theory(T), Exp(E) | Reference | $\Gamma(\pi^0 \to \gamma\gamma)$ eV | $\Gamma(\pi^0)$ eV | $\tau(\pi^0)$ ($10^{-17}$ s) |
|---|---|---|---|---|
| T (LO) | [50] | | 7.76 ± 0.02 | 8.48 ± 0.02 |
| T (HO) 2-loop ChPT | [56] | | 8.09 ± 0.11 | 8.14 ± 0.11 |
| E, PrimEx | [52] | 7.80 ± 0.12 | 7.90 ± 0.12 | 8.34 ± 0.13 |
| E, Direct | [53] | | 7.34 ± 0.28 | 8.97 ± 0.34 |
| E, $e^+e^-$ | [55] | 7.7 ± 0.7 | 7.8 ± 0.7 | 8.4 ± 0.8 |

Several high-order (HO) 3-flavor ChPT lifetime calculations have been conducted [50, 54, 56]. Since their results are approximately equal, we present only the two-loop HO calculation [56], alongside the 2-flavor LO theory results. The HO result for $\tau(\pi^0)$ is 4.0% lower than the LO value, which is attributed to HO corrections involving isospin breaking and the mixing of small η and η′ components into the $\pi^0$ wave function [30,54]. Additionally, the HO result is 2.4% lower than the PrimEx value. While PrimEx agrees better with 2-flavor LO results than with 3-flavor HO ChPT; both are consistent within the uncertainties. Overall, this supports the applicability of ChPT in the Chiral Anomaly sector. However, the uncertainties are too large to conclude definitively that the experimental results favor two-flavor ChPT. If future measurements confirm this with smaller uncertainties, it would indicate limitations in three-flavor ChPT.

In addition to the $\pi^0$ study, the JLab η lifetime study should offer more comprehensive tests of 3-flavor ChPT. JLab has already initiated studies of the Primakoff $\Gamma(\eta \to \gamma\gamma)$ decay width [57]. These data, along with further theoretical analyses, are essential for enhancing three-flavor ChPT tests.

## 10. The $\gamma \to 3\pi$ $F_{3\pi}$ Amplitude

In LO ChPT, $F_{3\pi} \approx e/(4\pi^2 f^3) = 9.72\ \mathrm{GeV}^{-3}$, where $e = \sqrt{4\pi\alpha_f} \approx 0.3028$, and f = 92.4 MeV is the pion decay constant [48,49]. Bijnens [58] examined higher-order ChPT corrections in the anomalous (intrinsic parity) sector, including one-loop diagrams having a WZW vertex and tree-level contributions from the $O(p^6)$ Lagrangian, with the $O(p^6)$ low-energy constants estimated using vector-meson-dominance (VMD). For $F_{3\pi}$, these HO corrections increased the LO value by ≈10%, resulting in a value estimated here as $F_{3\pi} \approx 10.7\ \mathrm{GeV}^{-3}$.

Antipov *et al.* at Serpukhov made the first measurement of $F_{3\pi}$ using a 40 GeV $\pi^-$ beam [59]. They used the Primakoff reaction $\pi^- Z \to \pi^-\pi^0 Z'$. The cross section depends on $(F_{3\pi})^2$. Their ≈200-event sample covered $-t < 2 \times 10^{-3}$ (GeV/c)$^2$ and $s(\pi^-\pi^0) < 10\ m_\pi^2$. They found $F_{3\pi} = 12.9 \pm 1.0$ GeV$^{-3}$, with statistical and systematic errors added here in quadrature. Ametller et al. [60], accounting for electromagnetic contributions, revised this value to $F_{3\pi} = 10.7 \pm 1.2$ GeV$^{-3}$. These data were also independently reanalyzed by Holstein [61], Hannah [62] and Truong [63]; who noted that incorporating momentum dependence reduces the experimental $F_{3\pi}$ value by ∼10%. Therefore, for the purposes of this review, the Serpukhov value is taken to be $F_{3\pi} = 9.6 \pm 1.2$ GeV$^{-3}$.

$F_{3\pi}$ was also measured through high-energy scattering at CERN involving pions interacting with electrons in $H_2$ molecular orbits: $\pi^- e \to \pi^- e \pi^0$. Amendolia *et al.* [64] reported 36 events for this reaction, corresponding to a cross-section of 2.11 ± 0.47 nb. Giller *et al.* [65] carried out Monte Carlo integrations of the cross-section within the kinematic range of the experiment using various theoretical expressions for $F_{3\pi}$. By employing an $O(p^6)$ SU(3) ChPT $F_{3\pi}$ amplitude, which including electromagnetic corrections, and accounting for the momentum transfer dependence of $F_{3\pi}$, they compared the integrated cross-section to the data and obtained $F_{3\pi} = 9.6 \pm 1.1$ GeV$^{-3}$. The quoted error reflects only the statistical uncertainty, not including any model dependence. The uncertainties in this determination are too large for precise ChPT testing. Recently, a dispersive analysis reported a value $F_{3\pi} \approx 10.0 \pm 0.5$ GeV$^{-3}$ based on $e^+e^- \to 3\pi$ data [66].

The Primakoff chiral anomaly $\pi\gamma \to \pi\pi$ and the radiative ρ production $\pi\gamma \to \rho \to \pi\pi$ reactions both lead to a $\pi^-\pi^0$ final state. COMPASS [67,68] carried out a high statistics measurement of the $\pi^-\pi^0$ Primakoff production cross-section from the kinematic threshold where the chiral anomaly dominates, through the region of the ρ(770) resonance. The main background comes from the high cross section diffractive Pomeron (ℙ) exchange reaction $\pi^-\mathbb{P} \to \pi^-\pi^0\pi^0$; when one $\pi^0$ has low energy and a second $\pi^0$ activates the trigger. This background was subtracted with the help of a Monte Carlo simulation. A potential background via the $\pi^-\mathbb{P} \to \pi^-\pi^0$ diffractive reaction is blocked by C-Parity. Since the Pomeron ℙ has the quantum numbers of the vacuum, it is even under charge conjugation (C = +1). Pomeron exchange can therefore only form a C eigenstate with eigenvalue C = +1. But the $|\pi^-\pi^0\rangle$ 2-pion state is not a C eigenstate with C = +1, since charge conjugation converts it to $|\pi^+\pi^0\rangle$. Therefore, C-parity forbids producing $|\pi^-\pi^0\rangle$.

Data analysis [67,68] extracted both the chiral anomaly amplitude $F_{3\pi}$ and the ρ radiative width $\Gamma(\rho \to \pi\gamma)$ using a dispersive theoretical approach [45]. This theory covers the full measured 0.3 – 1.1 Gev/c$^2$ $M_{\pi\pi}$ data range. It incorporates the ρ(770) by means of the ππ P-wave phase shift and the chiral anomaly calculated to 2-loop ChPT [62], including any interference between these two contributions. The number of Primakoff $\pi^-\pi^0$ events in each 0.5 GeV/c$^2$ mass bin was fitted to the theory. By using the 2.4% $K^-$ component of the COMPASS $\pi^-$ beam, it was possible to monitor the $\pi^-$ beam flux by measuring the number of $K^- \to \pi^-\pi^0$ decays in a decay volume outside of the target. Preliminary COMPASS results yield $F_{3\pi} = 10.3 \pm 0.1_{stat} \pm 0.6_{syst}$ GeV$^{-3}$ and $\Gamma(\rho \to \pi\gamma) \approx 76 \pm 9$ keV [67,68]. Ongoing analysis aims to reduce the backgrounds and systematic errors via cuts on additional kinematic variables, and to also include radiative corrections [67,68]. The preliminary $\Gamma(\rho\to \pi\gamma)$ value is consistent with the previous experimental results $\Gamma(\rho \to \pi\gamma) = (68 \pm 7)$ keV [33]. Table 2 summarizes the available experimental and theoretical $F_{3\pi}$ results.

Considering uncertainties, our discussion here focuses on the $F_{3\pi}$ amplitude most recently studied by COMPASS. Their preliminary value $F_{3\pi} = 10.3 \pm 0.6$ GeV$^{-3}$ lies about halfway between the 2-flavor LO ChPT prediction [37,45,65,69] and the 3-flavor ChPT prediction [58,69]. This value is consistent with both predictions, as well as with all other Table 2 results. A final COMPASS analysis with reduced uncertainty would enable improved testing of 3-flavor ChPT [70–72]. Proposed CERN AMBER studies of the chiral anomaly via the $\pi\gamma \to \pi\eta$ and $K\gamma \to K\pi^0$ Primakoff reactions, discussed below, along with further theoretical investigations, should provide additional input for testing of 3-flavor ChPT.

**Table 2.** Theoretical and experimental results for $\pi\gamma \to \pi\pi$, statistical and systematic errors are added in quadrature when both are given.

| Theory(T), Exp(E) | Reference | $F_{3\pi}$ GeV$^{-3}$ |
|---|---|---|
| Prim, COMPASS | [67,68] | 10.3 ± 0.6 |
| ChPT, LO T | [45,65,69] | 9.7 ± 0.1 |
| ChPT, HO T | [58] | 10.7 ± 1.0 |
| Prim, E + T Serpukhov | [59–63] | 9.6 ± 1.2 |
| E, CERN, $\pi e \to \pi e \pi^0$ | [64,65] | 9.6 ± 1.1 |
| E + T, | [66] | 10.0 ± 0.5 |

## 11. The $\pi\gamma \to \pi\eta$ $F_\eta$ amplitude

The $\pi\gamma \to \pi\eta$ CA Primakoff cross section depends on the square of the $F_\eta$ amplitude. The LO CA prediction is $F_\eta = F_{\eta\pi\pi} = e/(4\sqrt{3}\pi^2 f^3) = 5.61$ GeV$^{-3}$ [30,46], where $e = \sqrt{(4\pi\alpha)} \approx 0.3028$, and f = 92.4 MeV is the pion decay constant [48,49]. ChPT also predicts the dependence of this amplitude on the kinematic variables. Backgrounds to the $\pi^-\gamma \to \pi^-\eta$ CA Primakoff process come from Pomeron and Primakoff $a_2(1320)$ production. This reaction was studied by the VES collaboration with a 37 GeV/c $\pi^-$ beam and a beryllium target [73]. VES detected $\pi^-\pi^-\pi^+\gamma\gamma$, corresponding to $\eta \to \pi^-\pi^+\pi^0$; where the effective $\gamma\gamma$ mass equals the $\pi^0$ mass, and the effective mass of one of the $\pi^-\pi^+\pi^0$ subsystems equals the $\eta$ mass. The main background in the $\pi^+\pi^0\pi^-$ mass spectrum was from the low mass tail of $a_2(1320) \to \rho^0\pi^0 \to \pi^+\pi^-\pi^0$. Events in the range $M_{\pi\eta}$ = 0.70–1.18 GeV/c$^2$ were selected. After background subtraction, including from diffractive production of the $\pi^-\eta$ final state, N = 109 ± 23 events remained, corresponding to $F_\eta = 6.9 \pm 0.7$ GeV$^{-3}$. A dispersive formalism is available that includes both the CA Primakoff process as well as the Primakoff $a_2(1320)$ production processes [46]. When new high statistics data from CERN AMBER become available, this formalism may be used to deduce $F_\eta$.

## 12. The $K^-\gamma \to K^-\pi^0$ $F_{KK\pi}$ amplitude

The chiral anomaly amplitude $F_{KK\pi}$ is predicted to be equal to $F_{3\pi}$ at lowest order [47]. Compared to $\pi^-\gamma \to \pi^-\pi^0$, higher order chiral effects are expected to be more significant for $K^-\gamma \to K^0\pi^-$, considering the higher mass of kaons compared to pions. The $K^-\gamma \to K^-\pi^0$ and $K^-\gamma \to K^0\pi^-$ reactions are both possible for the $K^-\gamma \to K\pi$ Primakoff reaction. The chiral anomaly contributes to $K^-\gamma \to K^-\pi^0$, but not to the $K^-\gamma \to K^0\pi^-$ charge exchange reaction [47,74]. The main background is associated with Primakoff production of the $K^{*-}(892)$ resonance, and its subsequent decay to $K^-\pi^0$ or $K^0\pi^-$. In Ref. [74], $F_{KK\pi}$ is determined from the difference in cross section of the two $K^-\gamma \to K\pi$ processes near threshold. In Ref. [47], a dispersion theory treatment of the two $K^-\gamma \to K\pi$ reactions is described, based on modern $K\pi$ P-wave phase shift input. Dispersion theory relates the properties of a scattering amplitude to its singularities in the complex energy plane [75]. The amplitude is expressed in terms of its behavior at specific thresholds and branch points. Subtraction constants are introduced to handle divergences that arise in the calculation. They regularize the integrals that appear in the dispersion relations. A simultaneous fit to the two $K^-\gamma \to K\pi$ reactions would fix the subtraction constants of the theory, allowing $F_{KK\pi}$ to be determined. This chiral anomaly process may be studied as part of the planned CERN AMBER Primakoff experiments.

## 13. Theory Contributions to the Analysis of Primakoff Data

Primakoff reaction experiments have inspired ongoing theoretical research. For instance, in Ref. [76], a formalism is proposed to extract the $\gamma\pi \rightarrow \pi\pi$ chiral anomaly $F_{3\pi}$ from lattice QCD calculations conducted at pion masses larger than physical values.

In Ref. [45], a dispersive framework is established to extract both the $\gamma \rightarrow 3\pi$ $F_{3\pi}$ chiral anomaly amplitude (which contributes at low $\pi\pi$ mass) and the $\Gamma(\rho(770) \rightarrow \pi\gamma)$ radiative width (which strongly peaks at 770 MeV/$c^2$ $\pi\pi$ mass) from a fit to the Primakoff reaction $\gamma\pi \rightarrow \pi\pi$ cross-section data up to 1 GeV $\pi\pi$ mass. This analysis incorporates the physics of $\rho(770)$ through the $\pi\pi$ P-wave phase shift. In Ref. [77], this framework is extended and developed into a model-independent formalism that enables the extraction of $\Gamma(\rho \rightarrow \pi\gamma)$ directly from the residue of the resonance pole by analytically continuing the $\gamma\pi \rightarrow \pi\pi$ amplitude to the second Riemann sheet. These theoretical frameworks were integrated into the COMPASS data analysis.

In Ref. [78], a method is developed for extracting the kaon polarizabilities from $K\gamma \rightarrow K\gamma$ Primakoff scattering cross-sections. This approach utilizes dispersion theory to reconstruct the $K^*(892)$ contribution from its $K\pi$ intermediate state. Furthermore, to determine both $\alpha_K$ and $\beta_K$, the authors emphasize the necessity of measuring the $K\gamma$ Compton scattering angular distribution over a more complete angular range. They describe theoretical methods for analyzing kaon Primakoff data to extract kaon polarizabilities from $K\gamma \rightarrow K\gamma$ data, the chiral anomaly amplitude $F_{KK\pi}$ from $\gamma K \rightarrow K\pi^0$ data, and the $K^*(892) \rightarrow K\gamma$ radiative width.

## 14. Conclusions

Henry Primakoff pioneered scattering of high-energy beams from virtual photons in the Coulomb field of nuclei—the “Primakoff effect”—and proposed measuring the $\pi^0$ lifetime via $\gamma\gamma^*\rightarrow\pi^0$. This overview describes his personal life and career and reviews Primakoff experiments: pion polarizability and the $\pi\gamma\rightarrow\pi\pi$ anomaly at CERN COMPASS, the $\pi^0$ lifetime at JLab, and upcoming measurements (kaon polarizabilities, $\pi\gamma\rightarrow\pi\eta$, $K\gamma\rightarrow K\pi^0$) at CERN AMBER and an $\eta$ lifetime study at JLab. We note that ChPT agrees well with current results, supporting the pion’s pseudo-Goldstone status. We highlight how further Primakoff experiments and theoretical work will test and refine three-flavor (u,d,s) ChPT.

The transition from 2-flavor to 3-flavor ChPT incorporates the strange quark, which is essential for understanding the full dynamics of the light mesons (pions, kaons, and etas). Comparing 3-flavor ChPT predictions with kaon, $\pi^0$ and $\eta$ data should test how well the theory captures strange-quark effects and the pseudo-Goldstone dynamics. Any discrepancies would expose limitations of SU(3) ChPT and guide improvements or alternative approaches [70–72].

Low-energy Primakoff measurements and effective-Lagrangian comparisons complement high-energy perturbative QCD studies. Low-energy results constrain effective-theory parameters; high-energy data test short-distance inputs. Together they provide a unified check of QCD across scales and help refine both effective and perturbative descriptions.

This article was selected by Gist Science to be distributed to students, journalists, and curious non-specialists in ten languages in accessible explanation formats [79].

**Acknowledgments**

Thanks to Leonid Frankfurt, Bastian Kubis and Andrii Maltsev for their helpful comments, to Bakur Parsamyan (COMPASS) for providing Figure 3, and to Rafat Qubaja for valuable proofreading.

**ORCID**. Murray Moinester - https://orcid.org/0000-0001-8764-5618

**CRIS**. https://cris.tau.ac.il/en/persons/murray-moinester **Website**. https://murraymoinester.com